\documentclass[aps,prc,reprint,nofootinbib,amsmath]{revtex4-1}

\newcommand\I{\text{\boldmath$I$}}
\newcommand\ms[4]{\tfrac{#1}2\tfrac{#2}2\tfrac{#3}2\tfrac{#4}2}

\begin{document}

\title{Origin of a peculiar property of four particles in a $j=9/2$
  shell}

\author{K. Neerg\aa rd}

\affiliation{Fjordtoften 17, 4700 N\ae stved, Denmark}

\begin{abstract}
  It was found previously that in the system of four particles of the
  same kind in an angular momentum $j = 9/2$ shell, the energies $E_I$
  of four particular levels with angular momenta \linebreak $I = 4$,
  6, 10, and 12, are related by $E_I = a + b I(I+1)$, where $a$ and
  $b$ depend only on the two-body interaction. Here, I discuss
  observable implications of this regularity and analyse in detail its
  origin.
\end{abstract}

\maketitle

\section{\label{sec:intr}Introduction}

The single-$j$ shell model provides a crude description of low-energy
parts of the spectra of many nuclei. It may be applied, in particular,
by restriction to one kind of nucleon, to the spectra of some
semi-magic nuclei. A much studied example is the case of semi-magic
nuclei with four nucleons or four nucleon holes in an angular momentum
$j = 9/2$ shell, such as $^{72}$Ni and $^{74}$Ni with proton number
$Z = 28$ and four neutron or neutron holes in the $1g_{9/2}$ shell,
$^{94}$Ru and $^{96}$Pd with neutron number $N=50$ and four protons or
proton holes in the $1g_{9/2}$ shell, and $^{212}$Rn and $^{214}$Ra
with $N = 126$ and four protons or proton holes in the $2g_{9/2}$
shell. The angular momentum multiplets\footnote{The term
  \textit{angular momentum multiplet} is used to denote an irreducible
  module for the Lie algebra of angular momentum. A multiplet with
  angular momentum $I$ is thus $(2I+1)$-dimensional and has a basis of
  state vectors with magnetic quantum numbers
  \mbox{$M_I=I,I-1,\dots,-I$.}} of this system are grouped by angular
momentum $I$ and seniority $v$ as shown in Table~\ref{tbl:mult}, which
results from a straightforward count of combinations of
single-particle magnetic quantum numbers available to zero, two, and
four particles. In~\cite{ref:Nee22}, I define the subspace
$\Phi_4^\perp$ of the four-body state vector space of state vectors
$\chi$ that obey
\begin{equation}
  (a_m \phi)^{\dagger} a_{m'} \chi = 0
\end{equation}
for every state vector $\phi$ with angular momentum zero and every
pair of annihilators $a_m$ and $a_{m'}$ of particles with magnetic
quantum numbers $m$ and~$m'$. It follows from the analysis in
Sections~I and II of~\cite{ref:Nee22} that $\Phi_4^\perp$ is spanned
by four angular momentum multiplets, those with $I = 10$ and 12 and
one $v=4$ multiplet for each of $I=4$ and 6. The latter two, which I
call the \textit{Escuderos-Zamick} multiplets, are defined by having
zero coefficients of fractional parentage by the three-body multiplet
with $I = 9/2$ and $v = 3$. They are discussed extensively in the
literature; I refer to~\cite{ref:Qi26} for a review.\footnote{As
  detailed in my comment~\cite{ref:Nee26}, this review unfortunately
  accounts extremely inaccurately for the content of my
  article~\cite{ref:Nee22}.} Computer algebraic calculations reported
in~\cite{ref:Nee22} led to the discovery of the following
\textit{peculiar property of the four-body system}.

\textit{On $\Phi_4^\perp$, every rotationally invariant two-body
  interaction acts as a linear combination of the identity and the
  square of the total angular momentum.}

This regularity has observable implications to be discussed in
Section~\ref{sec:obs}. Retrospectively, the peculiar property could
have been inferred from previous computer algebraic results reported
in~\cite{ref:Isa08,ref:Qia18}. These authors write the two-body
interaction in the form
\begin{equation}\label{eq:VJ}
  V = \sum_J \nu_J V_J ,
\end{equation}
where $V_J$ projects on two-body angular momentum $J$. For $j =9/2$,
the two-body angular momentum $J$ takes the even values from 0 to $2j
-1 = 8$. Van Isacker and Heinze verify that the Escuderos-Zamick
multiplets are stationary for every $V$, that is, \textit{solvable} in
Talmi's terminology~\cite{ref:Tal10}, as discovered previously in
numeric calculations by Escuderos and Zamick~\cite{ref:Esc06}. They
find that the interaction energies $E_I$ in these multiplets are given
by $E_4 = \frac{68}{33} \nu_2 + \nu_4 + \frac{13}{15} \nu_6 +
\frac{114}{55} \nu_8$ and
$E_6 = \frac{19}{11} \nu_2 + \frac{12}{13} \nu_4 + \nu_6 +
\frac{336}{143} \nu_8$~\cite{ref:Isa08}. Qian and Qi consider the
levels with $I = 10$ and 12 and obtain
$E_{10} = \frac{23}{33} \nu_2 + \frac{98}{143} \nu_4 + \frac{233}{165}
\nu_6 + \frac{2292}{715} \nu_8$ and
$E_{12} = \frac{75}{143} \nu_4 + \frac{93}{55} \nu_6 + \frac{246}{65}
\nu_8$~\cite{ref:Qia18}.
\begin{table}
  \caption{\label{tbl:mult}Multiplicities of multiplets of four
    particles in a $j = 9/2$ shell per angular momentum $I$ and
    seniority $v$.}
\begin{ruledtabular}\begin{tabular}{c|cccccccccccc}
$v$&\hskip-15pt$I$\hskip1pt\vrule&0&2&3&4&5&6&7&8&9&10&12\\\hline
0&&1&0&0&0&0&0&0&0&0&0&0\\
2&&0&1&0&1&0&1&0&1&0&0&0\\
4&&1&1&1&2&1&2&1&1&1&1&1
\end{tabular}\end{ruledtabular}\end{table}           
Inspecting the coefficients $c_J^I$ in these expansions, one finds
that they obey $c_J^I = a_J + b_J I(I+1)$ with
$a_2,b_2 = \frac{26}{11},-\frac1{66 }$, \linebreak
$a_4,b_4 = \frac{153}{143},-\frac1{286}$,
$a_6,b_6 = \frac{41}{55},\frac1{165}$, and \linebreak
$a_8,b_8 =\frac{1308}{715},\frac9{715}$. The matrix $(c_J^I)$ thus has
rank~2. Since the pairing interaction $V_0$ kills $\Phi_4^\perp$ due
to the maximal seniority of this space, it follows that every $V$ acts
on $\Phi_4^\perp$ as a linear combination of two of them. Explicitly,
besides $V_0$, for example the interactions
$- 13 V_2 + 35 V_4 - 31 V_6 + 9 V_8$ and
$13 V_2 - 43 V_4 - 65 V_6 + 35 V_8$ kill $\Phi_4^\perp$. The two
interactions that provide by linear combination the action on
$\Phi_4^\perp$ of every $V$ may be taken as $n^2 - n$ and
$\I^2 - j(j+1) n$, where $n$ is the number operator, and $\I$ the
angular momentum vector. Since $n$ acts on four-body state vectors as
multiplication by the constant 4, the peculiar property follows. The
peculiar property implies that all angular momentum multiplets in
$\Phi_4^\perp$, and thus, in particular, the Escuderos-Zamick
multiplets, are solvable. Since the Escuderos-Zamick multiplets have
definite seniority, it follows that every $V$ conserves their
seniority. This has been described as a \textit{partial conservation
  of seniority}~\cite{ref:Isa08}.

After the discussion in Section~\ref{sec:obs} of observable
implications of the peculiar property, I analyse in detail, in
Sections~\ref{sec:obs}--\ref{sec:35t5h100}, how the cancellation of
$\Phi_4^\perp$ by the two aforesaid interactions comes about. Some
significant preconditions for this to happen are listed in the
concluding Section~\ref{sec:con}. Appendix~\ref{apx:proof} contains a
proof of validity of a parametrization of rotationally invariant
two-body interactions used in the analysis, and
Appendices~\ref{apx:j=9/2}--\ref{apx:35t5h100} provide some detailed
data referred to in the main part of the article.

\section{\label{sec:obs}Observable implications of the peculiar
property}

Throughout in this section, parts of the spectra to be discussed
are assumed to be well described by the $j = 9/2$ shell model. This
may not at all be taken as granted, given that the $j = 9/2$ shells are
not isolated but susceptible to perturbation by configurations
involving other shells. Experimental verification or falsification of
the following predictions thus serves as a test of the degree of
validity of the $j = 9/2$ shell model for the description of the
respective partial spectra. According to the $j = 9/2$ shell model,
the spectrum of a semi-magic nucleus with four particles or holes in a
$j = 9/2$ shell includes four parity even levels with angular momenta
$I = 4$, 6, 10, and~12, where those with $I = 4$ and 6 are
Escuderos-Zamick levels. By the peculiar property, the excitation
energies $E_I$ of these four levels are given by \linebreak
$E_I = a + b I(I+1)$ in terms of two constants $a$ and $b$. This
relation will be called the \textit{spectral rule}. The levels with
$I = 10$ and 12 are likely to be yrast, while the Escuderos-Zamick
levels may be non-yrast. If the $I = 6$ Escuderos-Zamick level is
below the yrast $I = 8$ level, which should have approximately
$v = 2$, it is expected to be strongly fed by $E2$ transition from
this level, which is otherwise predicted to be
isomeric~\cite{ref:Isa14}.

The nuclei $^{72}$Ni and $^{74}$Ni have a closed proton major shell
and four neutrons or neutron holes in the $1g_{9/2}$ shell. Their
spectra were studied by Morales~\textit{et
  al.}~\cite{ref:Mor16,ref:Mor18}. In both nuclei, two 4$^+$ and two
6$^+$ levels were observed with energies 1938 and 2323~keV for
$I = 4$, and 2392 and 2636~keV for $I = 6$, in $^{72}$Ni, and 1763 and
2105~keV for $I = 4$, and 2380 and 2507~keV for $I = 6$, in $^{74}$Ni.
All these assignments of spin and parity are tentative. Based on
observed transition probabilities, absence of isomerism of the yrast
8$^+$ level in $^{74}$Ni at 2605~keV, and large scale shell model
calculations, the authors of~\cite{ref:Mor18} suggest that the yrast
4$^+$ and 6$^+$ levels in $^{72}$Ni are the Escuderos-Zamick levels,
while in $^{74}$Ni, these are the second 4$^+$ and the yrast 6$^+$
levels. When this interpretation is adopted, extrapolation by the
spectral rule places the yrast 10$^+$ and 12$^+$ levels at 3795 and
4745~keV in $^{72}$Ni, and 3230 and 3805~keV in $^{74}$Ni. This seems
an unlikely difference between the yrast spectra of these two
neighboring even nuclei. Assuming in both nuclei the yrast 4$^+$ and
6$^+$ levels to be the Escuderos-Zamick levels, one gets 3795 and
4745~keV in $^{72}$Ni, and 4371 and 5717~keV in $^{74}$Ni, which is
also a fairly large difference. Assuming in both nuclei the
Escuderos-Zamick levels to be the second 4$^+$ and 6$^+$ levels, one
gets 3603 and 4258~keV in $^{72}$Ni, and 3750 and 4590~keV in
$^{74}$Ni.

The nuclei $^{94}$Ru and $^{96}$Pd have a closed neutron major shell
and four protons or proton holes in the $1g_{9/2}$ shell. Their
yrast 4$^+$, 6$^+$, 10$^+$, and 12$^+$ excitation energies are 2186.6,
2498.0, 3991.2, and 4716.6~keV in $^{94}$Ru, and 2099.01, 2424.19,
3783.6, and 4573.6~keV in $^{96}$Pd, with partly tentative assignments
of spin and parity~\cite{ref:NuDat}. Mach~\textit{et al.} and
Das~\textit{et al.} measured the $B(E2)$'s in the sequence of yrast
levels until $I = 8$~\cite{ref:Mac17,ref:Das22}. The $B(E2)$'s were
found consistent with $v = 0$ for $I = 0$, and $v = 2$ for $I =2$--8,
except for a strongly enhanced $B(E2,4^+ \rightarrow 2^+)$ in
$^{94}$Ru, which is explained by Das~\textit{et al.} to be due to
admixture of the 4$^+$ Escuderos-Zamick multiplet into the $v = 2$
multiplet by interaction with the $2p_{3/2}$ shell. Extrapolation
by the spectral rule from the observed 10$^+$ and 12$^+$ levels places
the pure Escuderos-Zamick levels at 2571.9 and 2918.9~keV in
$^{94}$Ru, and at 2237.9 and 2615.8~keV in $^{96}$Pd. In both cases
the 6$^+$ Escuderos-Zamick level is predicted above the yrast 8$^+$
level, which in $^{94}$Ru lies at 2644.1~keV and has a halflife of
71~$\mu$s, and in $^{96}$Pd at 2530.57~keV with a halflife of about
2.10~$\mu$s~\cite{ref:NuDat,ref:Mac17,ref:Das22}.

The nuclei $^{212}$Rn and $^{214}$Ra have a closed neutron major shell
and four protons or proton holes in the $2g_{9/2}$ shell. Their
yrast 4$^+$, 6$^+$, 10$^+$, and 12$^+$ excitation energies are
1501,42, 1639.68, 2654.9, and 2881.3~keV in $^{212}$Rn, and 1639.30,
1819.70, 2944.1, and 3329.4~keV in $^{214}$Ra. The nucleus $^{212}$Rn
has a second 4$^+$ level at 2324.43 keV, and tentatively assigned
6$^+$ levels at 2171.87, 2686.29, and 2865.47~keV. All the non-yrast
4$^+$ and 6$^+$ levels lie above the yrast 8$^+$ level at 1694.1~keV,
which is isomeric with a halflife of 0.91~$\mu$s, and none of the
6$^+$ levels feeds the latter~\cite{ref:NuDat}. Extrapolation by the
spectral rule from the observed 10$^+$ and 12$^+$ levels places the
Escuderos-Zamick levels at 2212 and 2320~keV. The first of these
energies agrees reasonably with the second observed 4$^+$ level while
the second one lies between the first and second tentatively assigned
non-yrast 6$^+$ levels. If the lowest tentatively assigned non-yrast
6$^+$ level would be the Escuderos-Zamick level, the order of the
Escuderos-Zamick levels would be opposite to the one predicted by the
spectral rule. In $^{214}$Ra, the spectral rule places the
Escuderos-Zamick levels at 2333 and 2482~keV, above the yrast 8$^+$
level at 1865.2~keV, which is isomeric with a halflife of
67.8~$\mu$s~\cite{ref:NuDat}.

\newpage

\section{\label{sec:int}Rotationally invariant two-body interaction}

\subsection{\label{sec:pre}Preliminaries}

A two-body interaction in a single-$j$ shell of one kind of particles
can be written
\begin{equation}\label{eq:Vn}
  V = \sum_{m_1>m_2,m_1'>m_2'}
    \langle m_1m_2|v_\text{n}|m_1'm_2' \rangle
    a_{m_1}^\dagger a_{m_2}^\dagger a_{m_2'} a_{m_1'}
\end{equation}
in terms of the usual annihilator $a_{m}$ of a particle with magnetic
quantum number $m$. It is rotationally invariant if and only if
$[I_0,V] = [I_\pm,V] = 0$, where
\begin{equation}
  I_\alpha = \sum_{m,m'} j_{\alpha,mm'} a_{m}^\dagger a_{m'}
\end{equation}
with
\begin{equation}\begin{aligned}
  & j_{0,mm'} = m \delta_{mm'} , \\
  & j_{\pm,mm'} = \sqrt{(j \mp m)(j \pm m + 1)} \delta_{mm'\pm1} .
\end{aligned}\end{equation}
The matrix $v_\text{n}$ is then a linear combination of terms with
entries
\begin{equation}\label{eq:cleb}
  \sum_{M=-J}^J \langle jm_1jm_2|JM \rangle
    \langle jm_1'jm_2'|JM\rangle ,
\end{equation}
where $J$ is even, and $\langle jm_1jm_2|JM \rangle$ is the
Clebsch-Gordan coefficient. So $v_\text{n}$ is symmetric. When this
symmetry is assumed, the relations $[I_\pm,V] = 0$ are equivalent, so
rotational invariance is equivalent to \linebreak
$[I_0,V] = [I_+,V] = 0$ and symmetry of $v_\text{n}$.

I define creators and annihilators
\begin{equation}\label{eq:alpha}
  \bar\alpha_m = \sqrt{\dfrac{(j+m)!}{(j-m)!}} \, a_m^\dagger , \quad
  \alpha_m = \sqrt{\dfrac{(j-m)!}{(j+m)!}} \, a_m ,
\end{equation}
associated with unnormalized one-body state
vectors.\footnote{In~\cite{ref:Nee22}, $\bar\alpha_m$ is called
  $\alpha_m^\dagger$. This notation is misleading because
  $\bar\alpha_m$ and $\alpha_m$ are not Hermitian conjugates.} The
advantage of these operators over the conventional $a_m^\dagger$ and
$a_m$ is that their commutators with $I_+$ take a simple form. Thus,
\begin{equation}\begin{gathered}\relax
  [I_0,\bar\alpha_m] = m \bar\alpha_m , \quad
  [I_+,\bar\alpha_m] = \bar\alpha_{m+1} , \\
  [I_-,\bar\alpha_m] = (j+m)(j-m+1) \bar\alpha_{m-1} , \\
  [I_0,\alpha_m] = - m \alpha_m , \quad
  [I_+,\alpha_m] = - \alpha_{m-1} , \\
  [I_-,\alpha_m] = - (j-m)(j+m+1) \alpha_{m+1} ,
\end{gathered}\end{equation}
except for $[I_+,\bar\alpha_j ]= [I_-,\bar\alpha_{-j}] =
[I_+,\alpha_{-j} ] = [I_-,\alpha_j] = 0$. The
definition~\eqref{eq:alpha} implies
\begin{equation}
  \{\bar\alpha_m,\alpha_{m'}\} = \delta_{m,m'} , \quad
   \bar\alpha_m = \dfrac{(j+m)!}{(j-m)!} \alpha_m^\dagger ,
\end{equation}
and~\eqref{eq:Vn} becomes
\begin{equation}
   V = \sum_{m_1>m_2,m_1'>m_2'}  \langle m_1m_2|v|m_1'm_2' \rangle
    \bar\alpha_{m_1} \bar\alpha_{m_2} \alpha_{m_2'} \alpha_{m_1'}
    \end{equation}
with 
\begin{multline}
  \langle m_1m_2|v|m_1'm_2' \rangle = \\
    \sqrt{\dfrac{(j-m_1)!(j-m_2)!(j+m_1')!(j+m_2')!}
      {(j+m_1)!(j+m_2)!(j-m_1')!(j-m_2')!}} \\
    \langle m_1m_2|v_\text{n}|m_1'm_2' \rangle .
\end{multline}
Symmetry of $v_\text{n}$ is equivalent to symmetry of $d v$, where the
matrix $d$ is diagonal with diagonal entries
\begin{equation}\label{eq:d(-M)}
  d_{m_1m_2} = \dfrac{(j+m_1)!(j+m_2)!}{(j-m_1)!(j-m_2)!} .
\end{equation}
The expression~\eqref{eq:cleb} implies
\begin{equation}
  \langle m_1m_2|v_\text{n}|m_1'm_2' \rangle
  = \langle -m'_2-m'_1|v_\text{n}|-m_2-m_1 \rangle ,
\end{equation}
whence
\begin{equation}\label{eq:sym}
  \langle m_1m_2|v|m_1'm_2' \rangle
  = \langle -m_2'-m_1'|v|-m_2-m_1  \rangle . 
\end{equation}

By $[I_0,V] = 0$, the matrix $v$ is block diagonal with blocks $v(M)$
connecting pairs $(m_1,m_2)$ and $(m'_1,m'_2)$ with
$m_1 + m_2 = m'_1 + m'_2 = M$. The pairs $(m_1,m_2)$ with
$m_1 + m_2 = M$ and $m_1 > m_2$ may be indexed in the order of
decreasing $m_1$ by an integer $i$ in the range
$0 \le i \le \lfloor i_M/2\rfloor$, where
\begin{equation}
  i_M = 2j -1 - |M| .
\end{equation}
Pairs $(m_1,m_2)$ and $(-m_2,-m_1)$ then have equal $i$. For
$M \ge 0$, one has $m_1 = j - i$, and for $M \le 0$, one has
\linebreak $m_2 = - j + i$. The relation~\eqref{eq:sym} becomes
\begin{equation}\label{eq:v(-M)}
  v(M) = v(-M)^T .
\end{equation}
When the diagonal matrix $d$ is split into blocks $d(M)$ corresponding
to the blocks $v(M)$, symmetry of $d v$ is equivalent to symmetry of
$d(M) v(M)$ for every $M$. The expression~\eqref{eq:d(-M)} implies
$d(M) = d(-M)^{-1}$. In particular, $d(0)$ is the identity matrix. It
is convenient to define also the diagonal matrix
\begin{equation}
  \bar d(M) = \frac{d(M)}{d(M)_{00}} .
\end{equation}
Symmetry of $d(M) v(M)$ is equivalent to symmetry of $\bar d(M) v(M)$.
The matrix $\bar d(M)$ has diagonal entries
\begin{equation}
  \frac{\dbinom{i_M+1}{i}\dbinom{2j-i_M+i-1}{i}}{\dbinom{2j}{i}} 
\end{equation}
for $M \ge 0$ and satisfies $\bar d(M) = \bar d(-M)^{-1}$. The matrix
$\bar d(0)$ is the identity matrix.

The relation $[I_+,V] = 0$ can be written \linebreak
$t(M) v(M-1) = v(M) t(M)$ for every $M$, where the matrix $t(M)$
connects $M$ and $M - 1$ and has entries
\begin{equation}\label{eq:t}
  t(M)_{ii'}  = \begin{cases} 1 , & i \le i' \le i + 1 , \\
      0 , & \text{otherwise} , \end{cases}
\end{equation}
for $M > 0$. For $M \le 0$, it is given by $t(M) = t(-M+1)^T$.

\subsection{\label{sec:par}Parametrization}

I use a parametrization of $V$ different from~\eqref{eq:VJ}. Its idea
is best illustrated by the relatively simple example of $j = 5/2$. In
this case, $|M| \le 4$. The $M = 4$ two-body space is 1-dimensional,
so one can write
\begin{equation}
  v(4) = (a)
\end{equation}
with arbitrary $a$. By $t(4) = (1)$, the relation
$t(4) v(3) = v(4) t(4)$ gives
\begin{equation}
  v(3) = (a) .
\end{equation}
By $t(3) = (1\ 1)$, the relation $t(3) v(2) = v(3) t(3)$ gives
\begin{equation}\label{eq:v2}
  v(2) = \begin{pmatrix} a - b & a - x \\ b & x \end{pmatrix}
\end{equation}
with arbitrary $b$ and $x$. Now,
\begin{equation}
  \bar d(2) = \begin{pmatrix} 1 & 0 \\ 0 & \frac95 \end{pmatrix} ,
\end{equation}
so for $\bar d(2) v(2)$ to be symmetric one must have
$a - x = \frac95 b$. Solving this equation with respect to $x$ and
inserting the solution in~\eqref{eq:v2} gives
\begin{equation}
  v(2) = \begin{pmatrix} a - b & \frac95 b \\
    b & a - \frac95 b \end{pmatrix} .
\end{equation}
By
\begin{equation}
  t(2) = \begin{pmatrix} 1 & 1 \\ 0 & 1 \end{pmatrix} ,
\end{equation}
the relation $t(2) v(1) = v(2) t(2)$ gives
\begin{equation}
  v(1) = \begin{pmatrix} a - 2 b & \frac85 b \\
    b & a - \frac45 b \end{pmatrix} .
\end{equation}
Finally, by
\begin{equation}
  t(1) = \begin{pmatrix} 1 & 1 & 0 \\ 0 & 1 & 1 \end{pmatrix} ,
\end{equation}
the relation $t(1) v(0) = v(1) t(1)$ gives
\begin{equation}\label{eq:v0}
  v(0) = \begin{pmatrix}
      a - 3 b + c & - \frac35 b + x & - a + \frac{12}5 b + y \\
      b  - c & a + \frac15 b - x & a - \frac45 b - y \\
      c & x & y \end{pmatrix} 
\end{equation}
with arbitrary $c$, $x$, and $y$. This matrix must be symmetric, whence
$- \frac35 b + x = b - c$ and $- a + \frac{12}5 b + y = c$. Solving
these equations with respect to $x$ and $y$ and inserting the solution
in~\eqref{eq:v0} gives
\begin{equation}
  v(0) = \begin{pmatrix}
      a - 3 b + c & b - c & c \\
      b  - c & a - \frac75 b + c & \frac85 b - c \\
      c & \frac85 b - c & a - \frac{12}5 b + c \end{pmatrix} .
\end{equation}
Now, \eqref{eq:v(-M)} gives $v(M)$ for $M<0$. The result is a
parametrization with parameters $a$, $b$, and $c$. When $V$ is written
in the form
\begin{equation}
  V = a V_a + b V_b + c V_c ,
\end{equation}
then $V_a$ is the identity on the two-body space, and $V_c$ is
proportional to the pairing interaction.

A similar expansion
\begin{equation}\label{eq:Vx}
  V = \sum_{i_x=0}^{j-\frac12} x_{i_x} V_{x_{i_x}}
\end{equation}
with $x_{i_x} = a, b, \dots$ for $i_x = 0, 1, \dots$\, exists for
every $j$. The interaction $V_{x_{i_x}}$ is given by
\begin{equation}\label{eq:vx}
  v_{x_{i_x}}(M) = w_{x_{i_x}}(M) \bar d(M) ,
\end{equation}
where $w_{x_{i_x}}(M)$ is defined, for $M \ge 0$, by
\begin{equation}\label{eq:w}
  w_{x_{i_x}}(M)_{ii'} = \sum_{i_d=0}^{\lfloor i_M/2\rfloor-i_x}
    c_{i_xi_Mi_d} u_{i_xi_Mi_d}(i) u_{i_xi_Mi_d}(i')
\end{equation}
with
\begin{equation}\label{eq:c}\begin{aligned}
  & c_{i_xi_Mi_d} = (-)^{i_x} \dbinom{2j-i_x}{i_d} \\
  & \hskip2em \biggl(
      \binom{i_M-i_x-2i_d}{i_x}\binom{i_M+1}{i_d}
      \binom{i_M-i_d+1}{i_d} \\
  & \hskip2em \binom{i_M-i_x-i_d+1}{i_d}
      \binom{2j-i_M+i_d-1}{i_d}
      \biggr)^{-1}
\end{aligned}\end{equation}
and
\begin{equation}\label{eq:u}\begin{aligned}
  & u_{i_xi_Mi_d}(i) = (-)^{i-i_d} \\
  & \hskip2em \dbinom{i}{i-i_d}\dbinom{i_M-i_x-i_d-i}{i_x+i_d-i}
    \dbinom{i_M-i+1}{i_d} .
  \end{aligned}\end{equation}
For $M < 0$, the matrix $w_{x_{i_x}}(M)$ is defined by \linebreak
$w_{x_{i_x}}(M) = \bar d(-M) w_{x_{i_x}}(-M) \bar d(-M) $. It is
easily verified that these definitions ensure symmetry of
$d v_{x_{i_x}}$, and~\eqref{eq:v(-M)} being satisfies by
$v_{x_{i_x}}(M)$.

The matrix $v_{x_{i_x}}(M)$ is zero unless $i_M \ge 2i_x$,
and $u_{i_xi_Mi_d}(i)$ is zero unless $i_d \le i \le i_d + i_x$. For
example, provided $j \ge \frac72$, the matrices
$v_c(2j - 1) = v_c(2j - 2) = v_c(2j - 3) = v_c(2j - 4) = 0$, and
\begin{multline}
  v_c(2j - 7) = \\ \hskip1em \begin{pmatrix}
     6 & -\frac{21(j-3)}j
       &  \frac{21(j-\frac52)(j-3)}{j(j-\frac12)}
       &  0 \\
    -3 &  \frac{12(j-\frac{11}4)}j
       & -\frac{18(j-\frac{13}6)(j-\frac52)}{j(j-\frac12)}
       &  \frac{15(j-2)(j-\frac52)}{j(j-\frac12)} \\
     1 & -\frac{6(j-\frac{13}6)}j
       &  \frac{16(j-\frac{23}{16})(j-\frac52)}{j(j-\frac12)}
       & -\frac{25(j-2)(j-\frac52)}{j(j-\frac12)} \\
    0  &  \frac{3(j-1)}j
       & -\frac{15(j-1)(j-\frac52)}{j(j-\frac12)}
       &  \frac{30(j-2)(j-\frac52)}{j(j-\frac12)}
  \end{pmatrix} .
\end{multline}
This is a sum of rank one, $3 \times 3$ matrices with $i_d = 0$ and 1
and upper left corners at $(i_d,i_d)$ with the counting from zero of
the matrix indices. In general, $v_{x_{i_x}}(M)_{ii'}$ is a rational
function of $j$ of equal degree $\min(i',i_x)$ in numerator and
denominator, and the denominator is
$\prod_{k=0}^{\min(i',i_x)-1} \linebreak (j - k \frac12)$. For every
$j$, one gets $v_a(M)_{ii'} = \delta_{ii'}$ and
$v_{x_{j-\frac12}}(M)_{ii'} = (-)^{j-\frac12+i+i'}\delta_{M0}$.
Rotational invariance of $V_{x_{i_x}}$ for every $j$, and every $i_x$
from 0 to $j - \frac12$, is proved in Appendix~\ref{apx:proof}, and
the matrices $v(M)$ for $j = 9/2$ are displayed in
Appendix~\ref{apx:j=9/2}.

Since the expansions~\eqref{eq:VJ} and~\eqref{eq:Vx} have equal
numbers of terms, and the interactions $V_{x_{i_x}}$ are linearly
independent, the parametrization~\eqref{eq:Vx} is complete. For $j =
9/2$, one finds
\begin{equation}
  \begin{pmatrix} V_0 \\ V_2 \\ V_4 \\ V_6 \\ V_8 \\ \end{pmatrix} =
  \begin{pmatrix}
    0 & 0 & 0 & 0 & \frac15 \\
    0 & 0 & 0 & -\frac1{11} & -\frac4{11} \\
    0 & 0 & \frac{18}{143} & \frac{27}{143} & \frac{162}{715} \\
    0 & -\frac3{10} & -\frac{12}{55} & -\frac7{55} & -\frac4{55} \\
    1 & \frac3{10} & \frac6{65} & \frac{21}{715} & \frac7{715}
  \end{pmatrix}
  \begin{pmatrix} V_a \\ V_b \\ V_c \\ V_d \\ V_e \\ \end{pmatrix}
\end{equation}
Hence,
$\frac1{12} (- 13 V_2 + 35 V_4 - 31 V_6 + 9 V_8) \equiv \linebreak
\frac34 V_a + V_b + V_c + V_d$ and
$\frac1{12} (13 V_2 - 43 V_4 - 65 V_6 + 35 V_8) \equiv
\frac{35}{12} V_a + \frac52 V_b + V_c$ modulo $V_e$. It is recalled
that these two interactions kill~$\Phi_4^\perp$.

\section{\label{seq:Phi}Four-body space}

From now on, $j = 9/2$. The four-body space $\Phi$ has a basis of
state vectors
\begin{equation}\label{eq:mmmm}
  |m_1m_2m_3m_4\rangle =
    \left( \prod_{i=1}^4 \bar a_{m_i} \right) |\rangle , 
\end{equation}
where $m_1>m_2>m_3>m_4$, and $|\rangle$ is an effective vacuum state
vector. In order to prove that a rotationally invariant interaction
$V$ kills the space $\Phi_4^\perp$, it is sufficient to prove that it
kills its subspace $\Phi_{40}^\perp$ of state vectors with $I_0 = 0$.
Therefore only the subspace $\Phi_{I_0=0}$ of $\Phi$ of state vectors
with $I_0 = 0$ needs to be considered. Moreover, since every angular
momentum eigenstate in $\Phi_{40}^\perp$ has even $I$, it is invariant
under the transformation
\begin{equation}
  R_y = \exp(-i \pi I_y) ,
\end{equation}
where $I_y = \frac1{2i} (I_+ - I_-)$. One can therefore further
diminish the space to be considered to the space
$\Phi_{I_0=0,\text e}$ of state vectors in $\Phi_{I_0=0}$ that are
even under the transformation $R_y$.

Since $R_y$ acts on the state vectors~\eqref{eq:mmmm} by
\begin{multline}\label{eq:Rym}
  R_y |m_1m_2m_3m_4\rangle = \\
    \left( \prod_{s=1}^4 \frac{(j+m_s)!}{(j-m_s)!} \right)
    |-m_4-m_3-m_2-m_1\rangle ,
\end{multline}
the space $\Phi_{I_0=0,\text e}$ has two kinds of basic state vectors,
state vectors $|m_1m_2-m_2-m_1\rangle$ with $m_1>m_2>0$, and state
vectors $(1 + R_y) |m_1m_2m_3m_4\rangle$ with $m_1+m_4 < 0$.
There are ten basic state vectors of the first kind, and four of the
second kind. The latter are
\begin{equation}\label{eq:omegal}\begin{aligned}
  \omega_1 = \, & |\ms{7}{5}{-3}{-9}\rangle
      + \frac73 |\ms{9}{3}{-5}{-7}\rangle := \\
    & |\ms{7}{5}{-3}{-9}\rangle
      + \alpha |\ms{9}{3}{-5}{-7}\rangle , \\
  \omega_2 = \, & |\ms{7}{3}{-1}{-9}\rangle
      + \frac83 |\ms{9}{1}{-3}{-7}\rangle := \\
    & |\ms{7}{3}{-1}{-9}\rangle
      + \beta |\ms{9}{1}{-3}{-7}\rangle , \\
  \omega_3 = \, & |\ms{5}{3}{1}{-9}\rangle
      + \frac{25}{6} |\ms{9}{-1}{-3}{-5}\rangle := \\
    & |\ms{5}{3}{1}{-9}\rangle
    + \gamma |\ms{9}{-1}{-3}{-5}\rangle1 , \\
  \omega_4 = \, & |\ms{5}{3}{-1}{-7}\rangle
      + \frac32 |\ms{7}{1}{-3}{-5}\rangle := \\
    & |\ms{5}{3}{-1}{-7}\rangle
      + \delta |\ms{7}{1}{-3}{-5}\rangle .
\end{aligned}\end{equation}
I denote these basic state vectors by $|r\rangle, r=1,\dots,14$. For
$r \le 10$, specifically $|r\rangle = |m_1m_2-m_2-m_1\rangle$ , where
the pairs $(m_1,m_2)$ with $m_1 > m_2 > 0$ are ordered lexically in
decreasing order of $m_1$ and $m_2$. For $r > 10$, specifically
$|r\rangle = \omega_{r-10}$. The coefficient of the second term in the
expression~\eqref{eq:omegal} for $\omega_l$ will be denoted by
$\xi_l$.
  
An arbitrary member of $\Phi_{I_0=0,\text e}$ can thus be written
\begin{equation}\label{eq:omega}
  \omega = \sum_{m_1>m_2>0} \kappa_{m_1m_2}
    |\langle m_1m_2-m_2-m_1 \rangle + \sum_{l=1}^4 \kappa_l \omega_l .
\end{equation}
I set $\kappa_{m_1m_2} = \kappa_{m_2m_1}$ for $m_1 < m_2$, and
$\kappa_{mm} = 0$, so that $(\kappa_{m_1m_2})$ becomes a symmetric
matrix. The entries of this matrix will be ordered by decreasing $m_1$
and $m_2$.

\section{\label{seq:Phi0}Angular momentum zero space}

The space $\Phi_0$ of four-body state vectors with angular momentum
zero is a subspace of $\Phi_{I_0=0,\text e}$. Its members $\phi$ are
characterized by $I_+ \phi = 0$. I express them in the
form~\eqref{eq:omega} by
\begin{equation}\label{eq:phi}
  \phi = \sum_{m_1>m_2>0} \zeta_{m_1m_2}
    |\langle m_1m_2-m_2-m_1 \rangle + \sum_{l=1}^4 \zeta_l \omega_l .
\end{equation}
Writing
\begin{multline}
  I_+ |m_1'm_2'm_3'm_4'\rangle = \\
  \hskip-5em \sum_{m_1>m_2>m_3>m_4} |m_1m_2m_3m_4\rangle \\
    \langle m_1m_2m_3m_4 |T| m_1'm_2'm_3'm_4' \rangle ,
\end{multline}
one has $\langle m_1m_2m_3m_4 |T| m_1'm_2'm_3'm_4' \rangle = 1$ when
\linebreak $m_i = m_i' + 1$ for some $i$, and $m_k = m_k'$ for
$k \ne i$. Otherwise,
$\langle m_1m_2m_3m_4 |T| m_1'm_2'm_3'm_4' \rangle = 0$. Hence, for
every state vector $|m_1m_2m_3m_4\rangle$ with $\sum_{i=1}^4 m_i = 1$,
the sum of coefficients in~\eqref{eq:phi} of state vectors
$|m_1'm_2'm_3'm_4'\rangle$ which obey this condition is zero. For 
$|m_1m_2m_3m_4\rangle = |\ms{7}{5}{-1}{-9}\rangle$, only
$|\ms{7}{5}{-3}{-9}\rangle$ and $|\ms{7}{3}{-1}{-9}\rangle$ obey the
condition, for $|m_1m_2m_3m_4\rangle = |\ms{7}{3}{1}{-9}\rangle$, only
$|\ms{7}{3}{-1}{-9}\rangle$ and $|\ms{5}{3}{1}{-9}\rangle$ obey the
condition, and for $|m_1m_2m_3m_4\rangle = |\ms{5}{3}{1}{-7}\rangle$,
only $|\ms{5}{3}{1}{-9}\rangle$ and $|\ms{5}{3}{-1}{-7}\rangle$ obey
the condition. Hence,
$\zeta_1 = - \zeta_2 = \zeta_3 = - \zeta_4 := q$.

For twelve state vectors $|m_1m_2m_3m_4\rangle$, two state vectors
$|m_1'm_2'm_3'm_4'\rangle$ with $m_1'+m_4'=0$ and either no, one or
two state vectors $|m_1'm_2'm_3'm_4'\rangle$ with $m_1'+m_4' \ne 0$
obey the condition. The resulting twelve relations imply
\newpage\noindent
\begin{equation}\label{eq:kappa}\begin{gathered}
  \zeta_{\frac92\frac72} + \zeta_{\frac92\frac52} = 0 , \quad
  \zeta_{\frac92\frac52} + \zeta_{\frac92\frac32} = -q , \\
  \zeta_{\frac92\frac32} + \zeta_{\frac92\frac12} = q , \quad
  \zeta_{\frac72\frac52} + \zeta_{\frac72\frac32} = -q \\
  \zeta_{\frac72\frac32} + \zeta_{\frac72\frac12} = 2q , \quad
  \zeta_{\frac52\frac32} + \zeta_{\frac52\frac12} = q , \\
  \zeta_{\frac92\frac52} + \zeta_{\frac72\frac52} = -\alpha q , \quad
  \zeta_{\frac92\frac32} + \zeta_{\frac72\frac32} =
    (-\alpha+\beta) q , \\
  \zeta_{\frac72\frac32} + \zeta_{\frac52\frac32} = \delta q , \quad
  \zeta_{\frac92\frac12} + \zeta_{\frac72\frac12} = \beta q , \\
  \zeta_{\frac72\frac12} + \zeta_{\frac52\frac12} = \delta q , \quad
  \zeta_{\frac52\frac12} + \zeta_{\frac32\frac12} = 0 .
\end{gathered}\end{equation}
Solving these equations gives
\vskip1pt\noindent
\begin{widetext}\begin{equation}\label{eq:zeta}
  (\zeta_{m_1m_2}) = \begin{pmatrix}
      0 & p & - p & p - q & - p + 2q \\
      p & 0 & p - \alpha q & - p + (-1+\alpha) q & p + (3-\alpha) q \\
      - p & p - \alpha q & 0  & p + (1-\alpha+\delta) q
        & - p + (\alpha-\delta) q \\
      p - q & - p + (-1+\alpha) q & p + (1-\alpha+\delta) q
        & 0 & p + (-\alpha+\delta) q \\
      - p + 2q & p + (3-\alpha) q & - p + (\alpha-\delta) q
        &  p + (-\alpha+\delta) q & 0
    \end{pmatrix} ,
  \end{equation}\end{widetext}
where $p$ is arbitrary. Consistency of the equations~\eqref{eq:kappa}
for $q \ne 0$ requires $-\alpha+\beta=-2+\alpha$, $\beta=5-\alpha$,
and $\delta=3-\delta$. Further, for
$|m_1m_2m_3m_4\rangle=|\ms{9}{1}{-3}{-5}\rangle$, three state vectors
$|m_1'm_2'm_3'm_4'\rangle$ with $m_1'+m_4' > 0$ obey the above
condition, giving $-\beta+\gamma-\delta=0$. These four equations
determine the constants $\xi_l$ independently of the symmetry
relation~\eqref{eq:Rym}.

The space $\Phi_0$ is spanned by the state vectors $\phi_0$ and
$\phi_1$ obtained by setting $p,q = 1,0$ and 0,1, respectively. The
state vectors $\phi_0$ and $3 \phi_0 + 4 \phi_1$ have $v = 0$ and 4,
respectively, \textit{cf.} Section III of~\cite{ref:Nee22}, where,
however, the choice of $\phi_1$ is different from the present one.
Alternatively, one may notice that state vectors of maximal seniority,
by carrying the representation $(1^n)$ of the symplectic group
$\text{Sp}(2j+1)$, are characterized by being \textit{traceless}, that
is, $\sum_{m'} (-)^{j-m'} \zeta_{mm'} = 0$ for every
$m$~\cite{ref:Wey39}. This is easily verified to hold for $p,q = 3,4$.
In the sequel, I denote by $\zeta^k_{m_1m_2}$ and $\zeta^k_l$ the
coefficients in the expansion~\eqref{eq:phi} of $\phi_k$.
\section{\label{seq:Phi40p}Space $\Phi_{40}^\perp$}

The state vectors $\chi \in \Phi_{40}^\perp$ are characterized by
$\phi^\dagger n_m \chi = 0$ for every $\phi \in \Phi_0$ and every $m$,
where $n_m = \bar \alpha_m \alpha_m$. I express them in the
form~\eqref{eq:omega} by
\begin{equation}\label{eq:chi}
  \chi = \sum_{m_1>m_2>0} \eta_{m_1m_2}
    |\langle m_1m_2-m_2-m_1 \rangle + \sum_{l=1}^4 \eta_l \omega_l .
\end{equation}
By the $R_y$-invariance, the relations $\phi^\dagger n_m \chi = 0$
with $\pm m$ are equivalent, so I assume $m > 0$. This gives ten
conditions to be satisfied by $\chi$, namely
\begin{equation}\label{eq:eqschi1}
  \phi_k^\dagger n_m \chi = 0
\end{equation}
for $k = 0$ and 1 and $m > 0$. This may be viewed as a system of
equations for the ten coefficients $\eta_{m_1m_2}$, given the
coefficients $\eta_l$. The matrix of this system of equations is
$(\kappa^{mk}_{m_1m_2})$ in terms of the coefficients
$\kappa^{mk}_{m_1m_2}$ in the expansion~\eqref{eq:omega} of
$n_m\phi_k$. Explicitly,
\begin{equation}
  \kappa^{mk}_{m_1m_2} =
    (\delta_{mm_1} + \delta_{mm_2}) \zeta^k_{m_1m_2} .
\end{equation}
These are simple fractions that are extracted easily
from~\eqref{eq:zeta}. The matrix $(\kappa^{mk}_{m_1m_2})$ is
non-singular, so every state vector in $\Phi_{40}^\perp$ is determined
by its coefficients~$\eta_l$, and $\Phi_{40}^\perp$ is spanned by the
state vectors $\chi_i$ with $\eta_l = \delta_{li}$.

By using
\begin{multline}
  \omega_{l}^\dagger n_m \omega_{l'} =
    \delta_{ll'} (1 - \delta_{m\mu_l})
    \prod_{s=1}^4 \frac{(j+m_s)!}{(j-m_s)!} = \\
    \delta_{ll'} (1 - \delta_{m\mu_l}) \xi_l ,
\end{multline}
where $\mu_l$ is the single $m > 0$ that does not occur in the
expression~\eqref{eq:omegal} for $\omega_l$, and the $m$'s are those
of the first term in this expression, in combination with \linebreak
$\zeta_l = (-)^{l-1} q$, one gets
\begin{equation}\label{eq:phinmom}
  \phi^\dagger n_m \omega_l = (-)^{l-1} (1 - \delta_{m\mu_l}) q \xi_l .
\end{equation}
I denote by $\eta^k_{m_1m_2}$ and $\eta^k_l$ the coefficients in the
expansion~\eqref{eq:chi} of~$\phi_k$. Equation~\eqref{eq:phinmom}
gives
\begin{equation}\label{eq:eqschi2}
  0 = \phi^\dagger n_m \chi_i = \sum_{m'} \zeta_{mm'} \eta^i_{mm'}
    + (-)^{i-1} (1 - \delta_{m\mu_i}) q \xi_i
\end{equation}
for every $\phi$, $m$, and $i$. One notices in passing that this
implies
\begin{equation}\label{eq:zeta0eta}
  \sum_{m'} \zeta^0_{mm'} \eta^i_{mm'} = 0
\end{equation}
for every $m$ and $i$. This says that every $\chi$ is traceless and
thus has $v = 4$.\footnote{The relation~\eqref{eq:zeta0eta} is, of
  course, nothing but $\chi_i$ being orthogonal to the $I_0 = 0$
  subspaces of the five multiplets with $v = 0$ and 2.}

\section{\label{seq:V}Action of $V$ on $\Phi_{I_0=0,\text{e}}$}
A matrix $(v_{rr'})$ may be defined by
\newpage\noindent
\begin{equation}
  V |r'\rangle = \sum_{r=1}^{14} |r\rangle v_{rr'} .
\end{equation}
For $r' \le 10$, I set $v_{rr'} = v^r_{m_1m_2}$, where \linebreak
$|m_1m_2-m_2-m_1\rangle = |r'\rangle$. Since $(\kappa^{mk}_{m_1m_2})$
is non-singular, another matrix $(\bar v^r_{mk})$ can be defined by
\begin{equation}\label{eq:barvrmk}
  v^r_{m_1m_2} = \sum_{mk} \bar v^r_{mk} \kappa^{mk}_{m_1m_2} .
\end{equation}
Then,
\begin{multline}\label{eq:sumveta}
  \sum_{m_1>m_2} v^r_{m_1m_2} \eta^i_{m_1m_2} =
    \sum_{mk} \bar v^r_{mk} \sum_{m_1>m_2} \kappa^{mk}_{m_1m_2}
    \eta^i_{m_1m_2} = \\
    \sum_{mk} \bar v^r_{mk}
    \sum_{m_1>m_2} (\delta_{mm_1} + \delta_{mm_2}) \zeta^k_{m_1m_2}
    \eta^i_{m_1m_2} = \\
    \sum_{mk} \bar v^r_{mk} \sum_{m'} \zeta^k_{mm'} \eta^i_{mm'} =
    (-)^i \xi_i \sum_m (1 - \delta_{m\mu_i}) \bar v^r_{m1}
\end{multline}
by~\eqref{eq:eqschi2}. For $V$ to kill $\Phi_{40}^\perp$, this must be
equal to $-v_{r,10+i}$ for every $r$ and $i$. One may notice that the
row vectors $(\bar v^r_{m0})$, which are proportional to the
contributions of $n_m \phi_0$ to the rows of $(v^r_{m_1m_1})$, do not
enter this criterion due to the maximal seniority of the state vectors
$\chi$. Only the contributions of $n_m \phi_1$ matter.

\section{\label{sec:3f1110}Interaction
  $V = \frac34 V_a + V_b + V_c + V_d$}
The matrices $v(M)$ of the interaction \linebreak 
$V = \frac34 V_a + V_b + V_c + V_d$ are shown in
Appendix~\ref{apx:3f1110}. \linebreak They give
\begin{widetext}\begin{equation}\label{eq:vrr'}
(v_{rr'})=\left(\begin{smallmatrix}
\tfrac{1}{2} &
-\tfrac{1}{3} &
-\tfrac{2}{3} &
\tfrac{8}{3} &
-2 &
1 &
0 &
0 &
0 &
0 &
\tfrac{14}{3} &
-\tfrac{16}{3} &
0 &
0 \\
-\tfrac{1}{3} &
\tfrac{1}{2} &
\tfrac{13}{6} &
-\tfrac{7}{6} &
2 &
0 &
0 &
1 &
0 &
0 &
-\tfrac{14}{3} &
0 &
-\tfrac{25}{3} &
0 \\
-\tfrac{2}{3} &
\tfrac{13}{6} &
-\tfrac{7}{2} &
\tfrac{7}{6} &
0 &
2 &
0 &
-2 &
0 &
0 &
\tfrac{14}{3} &
-\tfrac{16}{3} &
\tfrac{25}{3} &
0 \\
\tfrac{8}{3} &
-\tfrac{7}{6} &
\tfrac{7}{6} &
-\tfrac{15}{2} &
0 &
0 &
2 &
0 &
-2 &
1 &
0 &
\tfrac{16}{3} &
-\tfrac{25}{3} &
0 \\
-2 &
2 &
0 &
0 &
-\tfrac{53}{6} &
\tfrac{13}{6} &
-\tfrac{7}{6} &
-\tfrac{2}{3} &
\tfrac{8}{3} &
0 &
\tfrac{14}{3} &
0 &
0 &
-3 \\
1 &
0 &
2 &
0 &
\tfrac{13}{6} &
-\tfrac{29}{6} &
\tfrac{7}{6} &
-\tfrac{1}{3} &
0 &
\tfrac{8}{3} &
-\tfrac{14}{3} &
\tfrac{16}{3} &
0 &
3 \\
0 &
0 &
0 &
2 &
-\tfrac{7}{6} &
\tfrac{7}{6} &
\tfrac{19}{6} &
0 &
-\tfrac{1}{3} &
-\tfrac{2}{3} &
0 &
-\tfrac{16}{3} &
0 &
-3 \\
0 &
1 &
-2 &
0 &
-\tfrac{2}{3} &
-\tfrac{1}{3} &
0 &
\tfrac{7}{6} &
\tfrac{7}{6} &
-\tfrac{7}{6} &
\tfrac{14}{3} &
0 &
\tfrac{25}{3} &
-3 \\
0 &
0 &
0 &
-2 &
\tfrac{8}{3} &
0 &
-\tfrac{1}{3} &
\tfrac{7}{6} &
-\tfrac{17}{6} &
\tfrac{13}{6} &
0 &
0 &
-\tfrac{25}{3} &
3 \\
0 &
0 &
0 &
1 &
0 &
\tfrac{8}{3} &
-\tfrac{2}{3} &
-\tfrac{7}{6} &
\tfrac{13}{6} &
-\tfrac{17}{6} &
0 &
-\tfrac{16}{3} &
\tfrac{25}{3} &
-3 \\
1 &
-1 &
1 &
0 &
1 &
-1 &
0 &
1 &
0 &
0 &
0 &
0 &
0 &
0 \\
-1 &
0 &
-1 &
1 &
0 &
1 &
-1 &
0 &
0 &
-1 &
0 &
0 &
0 &
0 \\
0 &
-1 &
1 &
-1 &
0 &
0 &
0 &
1 &
-1 &
1 &
0 &
0 &
0 &
0 \\
0 &
0 &
0 &
0 &
-1 &
1 &
-1 &
-1 &
1 &
-1 &
0 &
0 &
0 &
0
\end{smallmatrix}\right) .
\end{equation}\end{widetext}
The part of $(v_{rr'})$ with $r$ or $r' \le 10$ is symmetric except
for an extra factor $2 \xi_{r'-10}$ for $r' > 10$. The off-diagonal
entries $v_{rr'}$ with $r,r' > 10$ (see next section) satisfy
$\xi_{r'-10} v_{rr'} = \xi_{r-10} v_{r'r}$. For $v_{rr'}$ to receive a
contribution from some $v(M)_{ii'}$, the state vectors $|r\rangle$ and
$|r'\rangle$ must share a pair of $m$'s. This strongly limits when it
can happen. The off-diagonal entries $v_{rr'}$ with $r,r' \le 10$ are
either equal to a single off-diagonal $v(0)_{ii'}$ or receive no
contribution at all. Otherwise, the off-diagonal entries are either
equal, for $r' \le 10$, or proportional to $\pm$ a single off-diagonal
$v(M)_{ii'}$ or receive no contribution at all, except for $r \le 10$
and $r' > 10$, when $v_{rr'}$ is the sum of two equal contributions or
none at all. The diagonal entries $v_{rr}$ are the sums of the
diagonal entries $v(M)_{ii}$ corresponding to the six pairs of $m$'s
in $|r\rangle$, or, for $r > 10$, in any one of the two terms in the
expression~\eqref{eq:omegal} for $|r\rangle$.

For the matrix $(\bar v^r_{m1})$ defined by~\eqref{eq:barvrmk}, one
gets
\begin{equation}
(\bar v^r_{m1})=\left(\begin{smallmatrix}
\tfrac{3}{2} &
\tfrac{3}{2} &
-\tfrac{1}{2} &
-\tfrac{1}{2} &
-\tfrac{1}{2} \\
-\tfrac{3}{2} &
\tfrac{1}{2} &
-\tfrac{3}{2} &
\tfrac{1}{2} &
\tfrac{1}{2} \\
\tfrac{3}{2} &
-\tfrac{1}{2} &
-\tfrac{1}{2} &
\tfrac{3}{2} &
-\tfrac{1}{2} \\
-\tfrac{3}{2} &
\tfrac{1}{2} &
\tfrac{1}{2} &
\tfrac{1}{2} &
-\tfrac{3}{2} \\
-\tfrac{1}{2} &
\tfrac{3}{2} &
\tfrac{3}{2} &
-\tfrac{1}{2} &
-\tfrac{1}{2} \\
\tfrac{1}{2} &
-\tfrac{3}{2} &
\tfrac{1}{2} &
-\tfrac{3}{2} &
\tfrac{1}{2} \\
-\tfrac{1}{2} &
\tfrac{3}{2} &
-\tfrac{1}{2} &
-\tfrac{1}{2} &
\tfrac{3}{2} \\
-\tfrac{1}{2} &
-\tfrac{1}{2} &
\tfrac{3}{2} &
\tfrac{3}{2} &
-\tfrac{1}{2} \\
\tfrac{1}{2} &
\tfrac{1}{2} &
-\tfrac{3}{2} &
\tfrac{1}{2} &
-\tfrac{3}{2} \\
-\tfrac{1}{2} &
-\tfrac{1}{2} &
-\tfrac{1}{2} &
\tfrac{3}{2} &
\tfrac{3}{2} \\
0 &
0 &
0 &
0 &
0 \\
0 &
0 &
0 &
0 &
0 \\
0 &
0 &
0 &
0 &
0 \\
0 &
0 &
0 &
0 &
0
\end{smallmatrix}\right) .
\end{equation}
\newpage\noindent
Right multiplication of this matrix by
\begin{equation}\label{eq:omd}
  \left(\begin{smallmatrix}
    1 & 1 & 1 & 0 & \\
    1 & 1 & 0 & 1 & \\
    1 & 0 & 1 & 1 & \\
    1 & 1 & 1 & 1 & \\
    0 & 1 & 1 & 1 &
  \end{smallmatrix}\right) ,
\end{equation}
representing the factor $1 - \delta_{m\mu_i}$ in~\eqref{eq:sumveta},
gives
\begin{equation}
\left(\begin{smallmatrix}
2 &
2 &
0 &
0 \\
-2 &
0 &
-2 &
0 \\
2 &
2 &
2 &
0 \\
0 &
-2 &
-2 &
0 \\
2 &
0 &
0 &
2 \\
-2 &
-2 &
0 &
-2 \\
0 &
2 &
0 &
2 \\
2 &
0 &
2 &
2 \\
0 &
0 &
-2 &
-2 \\
0 &
2 &
2 &
2 \\
0 &
0 &
0 &
0 \\
0 &
0 &
0 &
0 \\
0 &
0 &
0 &
0 \\
0 &
0 &
0 &
0
\end{smallmatrix}\right) .
\end{equation}
Multiplication of the $i$th column of this matrix by $(-)^{i-1}\xi_i$ is
seen to give the $(10+i)$th column of $(v_{rr'})$, so $V$
kills~$\Phi_{40}^\perp$.
\section{\label{sec:35t5h100}Interaction
  $V = \frac{35}{12} V_a + \frac52V_b + V_c$}
The case of the interaction $V =\frac{35}{12} V_a + \frac52V_b + V_c$
is similar with slightly less simple numbers. The matrices $v(M)$ are
shown in Appendix~\ref{apx:35t5h100}. They give
\begin{widetext}\begin{equation}
(v_{rr'})=\left(\begin{smallmatrix}
-\tfrac{1}{6} &
-\tfrac{4}{3} &
\tfrac{7}{3} &
0 &
1 &
0 &
0 &
0 &
0 &
0 &
-\tfrac{14}{3} &
0 &
0 &
0 \\
-\tfrac{4}{3} &
-\tfrac{5}{3} &
\tfrac{7}{6} &
\tfrac{7}{2} &
-\tfrac{5}{2} &
0 &
0 &
0 &
0 &
0 &
-\tfrac{7}{3} &
0 &
0 &
0 \\
\tfrac{7}{3} &
\tfrac{7}{6} &
-5 &
-\tfrac{5}{6} &
0 &
-\tfrac{5}{2} &
0 &
1 &
0 &
0 &
\tfrac{35}{3} &
\tfrac{8}{3} &
\tfrac{25}{3} &
0 \\
0 &
\tfrac{7}{2} &
-\tfrac{5}{6} &
-\tfrac{34}{3} &
0 &
0 &
-\tfrac{5}{2} &
0 &
1 &
0 &
0 &
\tfrac{40}{3} &
-\tfrac{25}{3} &
0 \\
1 &
-\tfrac{5}{2} &
0 &
0 &
-13 &
\tfrac{7}{6} &
\tfrac{7}{2} &
\tfrac{7}{3} &
0 &
0 &
\tfrac{35}{3} &
0 &
0 &
-7 \\
0 &
0 &
-\tfrac{5}{2} &
0 &
\tfrac{7}{6} &
-\tfrac{13}{3} &
-\tfrac{5}{6} &
-\tfrac{4}{3} &
0 &
0 &
-\tfrac{7}{3} &
\tfrac{40}{3} &
0 &
4 \\
0 &
0 &
0 &
-\tfrac{5}{2} &
\tfrac{7}{2} &
-\tfrac{5}{6} &
-\tfrac{10}{3} &
0 &
-\tfrac{4}{3} &
\tfrac{7}{3} &
0 &
\tfrac{8}{3} &
0 &
5 \\
0 &
0 &
1 &
0 &
\tfrac{7}{3} &
-\tfrac{4}{3} &
0 &
-\tfrac{23}{6} &
-\tfrac{5}{6} &
\tfrac{7}{2} &
-\tfrac{14}{3} &
0 &
\tfrac{25}{3} &
5 \\
0 &
0 &
0 &
1 &
0 &
0 &
-\tfrac{4}{3} &
-\tfrac{5}{6} &
-\tfrac{5}{6} &
\tfrac{7}{6} &
0 &
0 &
-\tfrac{25}{3} &
4 \\
0 &
0 &
0 &
0 &
0 &
0 &
\tfrac{7}{3} &
\tfrac{7}{2} &
\tfrac{7}{6} &
-\tfrac{7}{2} &
0 &
0 &
0 &
-7 \\
-1 &
-\tfrac{1}{2} &
\tfrac{5}{2} &
0 &
\tfrac{5}{2} &
-\tfrac{1}{2} &
0 &
-1 &
0 &
0 &
-\tfrac{28}{3} &
\tfrac{8}{3} &
-\tfrac{25}{6} &
\tfrac{3}{2} \\
0 &
0 &
\tfrac{1}{2} &
\tfrac{5}{2} &
0 &
\tfrac{5}{2} &
\tfrac{1}{2} &
0 &
0 &
0 &
\tfrac{7}{3} &
-\tfrac{32}{3} &
\tfrac{25}{6} &
-\tfrac{3}{2} \\
0 &
0 &
1 &
-1 &
0 &
0 &
0 &
1 &
-1 &
0 &
-\tfrac{7}{3} &
\tfrac{8}{3} &
-\tfrac{50}{3} &
\tfrac{3}{2} \\
0 &
0 &
0 &
0 &
-\tfrac{7}{3} &
\tfrac{4}{3} &
\tfrac{5}{3} &
\tfrac{5}{3} &
\tfrac{4}{3} &
-\tfrac{7}{3} &
\tfrac{7}{3} &
-\tfrac{8}{3} &
\tfrac{25}{6} &
-6
\end{smallmatrix}\right) .
\end{equation}\end{widetext}
Hence,
\begin{equation}
(\bar v^r_{m1})=\left(\begin{smallmatrix}
-\tfrac{1}{2} &
-\tfrac{1}{2} &
-\tfrac{1}{2} &
-\tfrac{1}{2} &
\tfrac{3}{2} \\
\tfrac{1}{2} &
\tfrac{1}{2} &
\tfrac{1}{2} &
-\tfrac{5}{2} &
\tfrac{3}{2} \\
\tfrac{3}{2} &
-\tfrac{1}{2} &
\tfrac{5}{2} &
\tfrac{3}{2} &
-\tfrac{7}{2} \\
-\tfrac{7}{2} &
-\tfrac{3}{2} &
\tfrac{3}{2} &
\tfrac{7}{2} &
-\tfrac{7}{2} \\
-\tfrac{1}{2} &
\tfrac{25}{6} &
\tfrac{25}{6} &
-\tfrac{17}{6} &
-\tfrac{5}{6} \\
\tfrac{1}{2} &
-\tfrac{13}{6} &
\tfrac{17}{6} &
-\tfrac{13}{6} &
-\tfrac{7}{6} \\
\tfrac{3}{2} &
-\tfrac{11}{6} &
-\tfrac{5}{6} &
\tfrac{7}{6} &
-\tfrac{11}{6} \\
\tfrac{5}{2} &
-\tfrac{17}{6} &
-\tfrac{5}{6} &
-\tfrac{5}{6} &
\tfrac{7}{6} \\
\tfrac{3}{2} &
\tfrac{5}{6} &
-\tfrac{7}{6} &
-\tfrac{7}{6} &
-\tfrac{7}{6} \\
-\tfrac{7}{2} &
\tfrac{7}{6} &
\tfrac{7}{6} &
\tfrac{7}{6} &
\tfrac{7}{6} \\
-1 &
-1 &
-1 &
-1 &
2 \\
1 &
1 &
-2 &
1 &
1 \\
-1 &
2 &
-1 &
-1 &
-1 \\
-2 &
1 &
1 &
1 &
1
\end{smallmatrix}\right) .
\end{equation}
Right multiplication of this matrix by the matrix~\eqref{eq:omd} gives
\begin{equation}
\left(\begin{smallmatrix}
-2 &
0 &
0 &
0 \\
-1 &
0 &
0 &
0 \\
5 &
-1 &
2 &
0 \\
0 &
-5 &
-2 &
0 \\
5 &
0 &
0 &
\tfrac{14}{3} \\
-1 &
-5 &
0 &
-\tfrac{8}{3} \\
0 &
-1 &
0 &
-\tfrac{10}{3} \\
-2 &
0 &
2 &
-\tfrac{10}{3} \\
0 &
0 &
-2 &
-\tfrac{8}{3} \\
0 &
0 &
0 &
\tfrac{14}{3} \\
-4 &
-1 &
-1 &
-1 \\
1 &
4 &
1 &
1 \\
-1 &
-1 &
-4 &
-1 \\
1 &
1 &
1 &
4
\end{smallmatrix}\right) .
\end{equation}
\newpage\noindent
Multiplication of the $i$th column of this matrix by $(-)^{i-1}\xi_i$ is
seen to give the $(10+i)$th column of $(v_{rr'})$, so $V$
kills~$\Phi_{40}^\perp$.

\section{\label{sec:con}Conclusion}

In conclusion, the peculiar property of four particles in an angular
momentum $j=9/2$ shell that every rotationally invariant two-body
interaction acts as a linear combination of the identity and the
square of the total angular momentum, on the space $\Phi_4^\perp$
spanned by the two Escuderos-Zamick angular momentum multiplets and
the multiplets with $I = 10$ and 12, is due to the fact that a
three-dimensional vector space of such interactions kills
$\Phi_4^\perp$. This space of interactions is spanned by the pairing
interaction, which kills $\Phi_4^\perp$ due to the maximal seniority
of the latter, and two other interactions which have simple parameters
in the parametrization introduced in Section~\ref{sec:par}. Several
specific features of the system are significant for the cancellation
of $\Phi_4^\perp$ by the latter two. (\textit{i})~The space
$\Phi_4^\perp$ is the orthogonal complement of the span of the set of
state vectors obtained by application of an arbitrary one-body
operator to a state vector with angular momentum zero.
(\textit{ii})~The system has two multiplets with angular momentum
zero, which have seniorities 0 and 4. (\textit{iii})~The space of
state vectors with magnetic quantum number zero that are even under
half-turn rotation about the $y$-axis has a basis consisting of ten
state vectors composed of two pairs of single-particle state vectors
with opposite magnetic quantum numbers in each pair, and four linear
combinations of two state vectors composed of single-particle state
vectors with different absolute values of their magnetic quantum
numbers. The number of basic state vectors of the first kind equals
the dimension of the zero angular momentum space times half the
dimension of the single-particle state vector space, while the number
of basic state vectors of the second kind equals the number of angular
momentum multiplets in $\Phi_4^\perp$. (\textit{iv})~The number of
particles is one less than half the dimension of the single-particle
state vector space.

\appendix

\section{\label{apx:proof}Rotational invariance of $V_{x_{i_x}}$}

By construction, $[I_0,V_{x_{i_x}}] = 0$, and $d v_{x_{i_x}}$ is
symmetric. To prove that $V_{x_{i_x}}$ is rotationally invariant, it
remains to prove $[I_+,V_{x_{i_x}}] = 0$. As noted at the end of
Section~\ref{sec:pre}, this relation can be written
$t(M) v_{x_{i_x}}(M-1) = v(M)_{x_{i_x}} t(M)$ for every $M$. Since
$v_{x_{i_x}}(M)$ satisfies~\eqref{eq:v(-M)} by construction, and
$t(M) = t(-M+1)^T$, it is sufficient to prove
$t(M) v_{x_{i_x}}(M-1) = v(M)_{x_{i_x}} t(M)$ for $M > 0$.
By~\eqref{eq:vx}, the relation can then be written
\begin{equation}\label{eq:tw=wt}
  t(M) w_{x_{i_x}}(M-1) = w_{x_{i_x}}(M) \bar t(M)
\end{equation}
with $\bar t(M) = \bar d(M)\ t(M)\bar d(M-1)^{-1}$. Explicitly,
\begin{multline}\label{eq:tbar}
  \langle i|\bar t(M)|i' \rangle = \\
  \begin{cases} \dfrac{(iM-i+2)(2j-i_M+i-1)}{(i_M+2)(2j-i_M-1)} ,
      & i' = i , \\[10pt]
    \dfrac{(i+1)(2j-i)}{(i_M+2)(2j-i_M-1)} , & i' = i + 1 , \\[10pt]
      0 , & \text{otherwise} . \end{cases}
\end{multline}
When the column vector with entries $u_{i_xi_Mi_d}(i)$ is denoted by
$u_{i_xi_Mi_d}$, the relation \eqref{eq:tw=wt} takes, by~\eqref{eq:w},
the form
\begin{multline}
  \sum_{i_d=0}^{\lfloor (i_M+1)/2\rfloor-i_x}
    t(M) u_{i_xi_M+1i_d} c_{i_xi_M+1i_d} u_{i_xi_M+1i_d}^T = \\
  \sum_{i_d=0}^{\lfloor i_M/2\rfloor-i_x}
    u_{i_xi_Mi_d} c_{i_xi_Mi_d}  u_{i_xi_Mi_d}^T \bar t(M) .
\end{multline}

Now assume that there exist numbers $a_{i_xi_Mi_d}$ and
$b_{i_xi_Mi_d}$ such that
\begin{equation}\label{eq:ab}
  t(M) u_{i_xi_M+1i_d}
  = a_{i_xi_Mi_d} u_{i_xi_Mi_d} + b_{i_xi_Mi_d} u_{i_xi_Mi_d-1}
\end{equation}
with omission of the first term when $i_M$ is odd and \linebreak
$i_d = \lfloor (i_M+1)/2\rfloor-i_x$, and the second term when
$i_d = 0$. Then
\begin{multline}\label{eq:comab}
  \sum_{i_d=0}^{\lfloor (i_M+1)/2\rfloor-i_x}
      t(M) u_{i_xi_M+1i_d} c_{i_xi_M+1i_d} u_{i_xi_M+1i_d}^T = \\
  \sum_{i_d=0}^{\lfloor (i_M+1)/2\rfloor-i_x}
      ( a_{i_xi_Mi_d} u_{i_xi_Mi_d}
        + b_{i_xi_Mi_d} u_{i_xi_Mi_d-1} ) \\
      c_{i_xi_M+1i_d} u_{i_xi_M+1i_d}^T .
\end{multline}
Similarly, if there exist numbers $e_{i_xi_Mi_d}$ and
$f_{i_xi_Mi_d}$ such that
\begin{multline}\label{eq:ef}
  u_{i_xi_Mi_d}^T \bar t(M) = \\
  e_{i_xi_Mi_d} u_{i_xi_M+1i_d}^T + f_{i_xi_Mi_d}
  u_{i_xi_M+1i_d+1}^T
\end{multline}
with omission of the second term when $i_M$ is even and
$i_d = \lfloor i_M/2\rfloor - i_x$, then
\begin{multline}\label{eq:comef}
  \sum_{i_d=0}^{\lfloor i_M/2\rfloor-i_x}
    u_{i_xi_Mi_d} c_{i_xi_Mi_d}  u_{i_xi_Mi_d}^T \bar t(M) = \\
 \hskip-10em \sum_{i_d=0}^{\lfloor i_M/2\rfloor-i_x}
   u_{i_xi_Mi_d} c_{i_xi_Mi_d} \\
   ( e_{i_xi_Mi_d} u_{i_xi_M+1i_d}^T
     + f_{i_xi_Mi_d} u_{i_xi_M+1i_d+1}^T ).
\end{multline}
The right hand sides of~\eqref{eq:comab} and~\eqref{eq:comef} are
equal when
\begin{equation}\label{eq:ab=ef}\begin{gathered}
  a_{i_xi_Mi_d} c_{i_xi_M+1i_d} = c_{i_xi_Mi_d} e_{i_xi_Mi_d} , \\
  b_{i_xi_Mi_d+1} c_{i_xi_M+1i_d+1} = c_{i_xi_Mi_d} f_{i_xi_Mi_d} ,
\end{gathered}\end{equation}
for every pair of $a_{i_xi_Mi_d}$ and $e_{i_xi_Mi_d}$, and of
$b_{i_xi_Mi_d+1}$ and $f_{i_xi_Mi_d}$, in~\eqref{eq:ab}
and~\eqref{eq:ef}.

By~\eqref{eq:t}, Equation~\eqref{eq:ab} can be written
\begin{multline}\label{eq:abi}
  u_{i_xi_M+1i_d}(i) + u_{i_xi_M+1i_d}(i+1) = \\
  a _{i_xi_Mi_d} u_{i_xi_Mi_d}(i)
  + b _{i_xi_Mi_d} u_{i_xi_Mi_d-1}(i) .
\end{multline}
This remains true when $i + 1$ exceeds the range of vector indices of
$u_{i_xi_M+1i_d}$, which happens when $iM$ is even and
$i = \lfloor i_M/2\rfloor$, because then $u_{i_xi_M+1i_d}(i+1) = 0$
by~\eqref{eq:u}. When $u_{i_xi_M+1i_d}(i) \ne 0$, \eqref{eq:abi} can
be written
\begin{multline}\label{eq:1+r=as+bt}
  1 + \frac{u_{i_xi_M+1i_d}(i+1)}{u_{i_xi_M+1i_d}(i)} = \\
    \hskip1em a _{i_xi_Mi_d}
      \frac{u_{i_xi_Mi_d}(i)}{u_{i_xi_M+1i_d}(i)}
    + b _{i_xi_Mi_d} \frac{u_{i_xi_Mi_d-1}(i)}{u_{i_xi_M+1i_d}(i)} .
  \end{multline}    
Equation~\eqref{eq:u} gives
\begin{equation}\begin{aligned}
  & \frac{u_{i_xi_M+1i_d}(i+1)}{u_{i_xi_M+1i_d}(i)} = \\
  & \hskip1em \frac{(i+1)(i_x+i_d-i)(i_M-i_d-i+2)}
        {(i_d-i-1)(i_M-i_x-i_d-i+1)(i_M-i+2)} , \\
  & \frac{u_{i_xi_Mi_d}(i)}{u_{i_xi_M+1i_d}(i)} = \\
  & \hskip1em \frac{(i_M-2i_x-2i_d+1)(i_M-i_d-i+2)}
    {(i_M-i_x-i_d-i+1)(i_M-i+2)} , \\
  & \frac{u_{i_xi_Mi_d-1}(i)}{u_{i_xi_M+1i_d}(i)} = \\
  & \hskip1em \frac{i_d^2(i_x+i_d-i)}
      {(i_d-i-1)(i_M-2i_x-2i_d+2)(i_M-i+2)} .
\end{aligned}\end{equation}
When these expressions and
\begin{equation}\label{eq:a,b}\begin{gathered}
  a _{i_xi_Mi_d} =  \frac{i_M-i_x-i_d+2}{i_M-i_x-2i_d+2} , \\
  b _{i_xi_Mi_d} =
    \frac{(i_M-2i_d+3)(i_M-2i_x-2i_d+2)}{i_d(i_M-i_x-2i_d+2)} ,
\end{gathered}\end{equation}
are inserted in~\eqref{eq:1+r=as+bt}, it becomes
\newpage\noindent
\begin{multline}\label{eq:verab}
  1 + \frac{(i+1)(i_x+i_d-i)(i_M-i_d-i+2)}
      {(i_d-i-1)(i_M-i_x-i_d-i+1)(i_M-i+2)}  = \\
    \frac{(i_M-i_x-i_d+2)(i_M-2i_x-2i_d+1)(i_M-i_d-i+2)}
      {(i_M-i_x-2i_d+2)(i_M-i_x-i_d-i+1)(i_M-i+2)} \\
    + \frac{(i_M-2i_d+3)i_d(i_x+i_d-i)}
      {(i_M-i_x-2i_d+2)(i_d-i-1)(i_M-i+2)} .
    \end{multline}
This relation holds identically, so~\eqref{eq:ab} is true for
$u_{i_xi_M+1i_d}(i) \ne 0$ with $a _{i_xi_Mi_d}$ and $b _{i_xi_Mi_d}$
given by~\eqref{eq:a,b}. The algebraic identity~\eqref{eq:verab}
remains true upon extension to real $i$. By
multiplying~\eqref{eq:verab} by $u_{i_xi_M+1i_d}(i)$ and taking the
appropriate limits, one then gets~\eqref{eq:abi} for
$u_{i_xi_M+1i_d}(i) = 0$. Thus~\eqref{eq:ab} holds with
$a _{i_xi_Mi_d}$ and $b _{i_xi_Mi_d}$ given by~\eqref{eq:a,b}. The
omission of the first term in~\eqref{eq:ab} when $i_M$ is odd and
$i_d = \lfloor (i_M+1)/2\rfloor-i_x$ does not invalidate this
conclusion because the first term on the right hand side
of~\eqref{eq:verab} vanishes in this case. Neither does the omission
of the second term in~\eqref{eq:ab} when $i_d = 0$ invalidate the
conclusion because the second term on the right hand side
of~\eqref{eq:verab} vanishes in this case.

Similarly, by~\eqref{eq:tbar}, Equation~\eqref{eq:ef} can be written
\begin{equation}\label{eq:efi}\begin{aligned}
  & \frac{(i_M-i+2)(2j-i_M+i-1)}{(i_M+2)(2j-i_M-1)}
      u_{i_xi_Mi_d}(i) \\
  & \hskip1em + \frac{i(2j-i+1)}{(i_M+2)(2j-i_M-1)}
      u_{i_xi_Mi_d}(i-1) = \\
  & \hskip1em e _{i_xi_Mi_d} u_{i_xi_M+1i_d}(i)
      + f _{i_xi_Mi_d} u_{i_xi_M+1i_d+1}(i).
\end{aligned}\end{equation}
Like before, this remains true for $i = 0$. When
$u_{i_xi_Mi_d}(i) \ne 0$, \eqref{eq:efi} is equivalent to
\begin{equation}\label{eq:p+qr=as+bt}\begin{aligned}
  & \frac{(i_M-i+2)(2j-i_M+i-1)}{(i_M+2)(2j-i_M-1)} \\
  & \hskip1em + \frac{i(2j-i+1)}{(i_M+2)(2j-i_M-1)}
  \frac{u_{i_xi_Mi_d}(i-1)}{u_{i_xi_Mi_d}(i)} = \\
  & \hskip1em e _{i_xi_Mi_d}
    \frac{u_{i_xi_M+1i_d}(i)}{u_{i_xi_Mi_d}(i)}
     + f _{i_xi_Mi_d} \frac{u_{i_xi_M+1i_d+1}(i)}{u_{i_xi_Mi_d}(i)} .
\end{aligned}\end{equation}
Equation~\eqref{eq:c} gives\newpage
\begin{widetext}\begin{equation}\label{eq:d/d}\begin{gathered}
  \frac{c_{i_xi_M+1i_d}}{c_{i_xi_Mi_d}} = \frac{(i_M-2i_x-2i_d+1)
      (i_M-2i_d+2)(i_M-i_x-2i_d+2)(2j-i_M+i_d-1)}
    {(i_M-i_x-2i_d+1)(i_M+2)(i_M-i_x-i_d+2)(2j-i_M-1)} , \\
  \frac{c_{i_xi_M+1i_d+1}}{c_{i_xi_Mi_d}} =
    \frac{(2j-i_x-i_d)(i_M-i_x-2i_d)(i_d+1)^3}
      {(i_M-2i_x-2i_d)(i_M+2)(i_M-2i_d+1)(i_M-i_x-2i_d+1)(2j-i_M-1)} .
\end{gathered}\end{equation}\end{widetext}
By~\eqref{eq:a,b} and~\eqref{eq:d/d}, the relations~\eqref{eq:ab=ef}
are satisfied when
\begin{equation}\label{eq:e,f}\begin{aligned}
    & e _{i_xi_Mi_d} =
       \frac{c_{i_xi_M+1i_d}}{c_{i_xi_Mi_d}} a_{i_xi_Mi_d} = \\
  & \hskip1em \frac{(i_M-2i_x-2i_d+1)(i_M-2i_d+2)(2j-i_M+i_d-1)}
      {(i_M-i_x-2i_d+1)(i_M+2)(2j-i_M-1)} , \\
  & f _{i_xi_Mi_d} =
        \frac{c_{i_xi_M+1i_d+1}}{c_{i_xi_Mi_d}} b_{i_xi_Mi_d+1} =  \\
  & \hskip1em \frac{(2j-i_x-i_d)(i_d+1)^2}
      {(i_M+2)(i_M-i_x-2i_d+1)(2j-i_M-1)} .
    \end{aligned}\end{equation}
\newpage\noindent
Equation~\eqref{eq:u} gives
\begin{equation}\label{eq:u/u}\begin{aligned}
  & \frac{u_{i_xi_Mi_d}(i-1)}{u_{i_xi_Mi_d}(i)} = \\
  & \hskip1em \frac{(i_d-i)(i_M-i_x-i_d-i+1)(i_M-i+2)}
      {i(i_x+i_d-i+1)(i_M-i_d-i+2)} , \\
  & \frac{u_{i_xi_M+1i_d}(i)}{u_{i_xi_Mi_d}(i)} = \\
  & \hskip1em \frac{(i_M-i_x-i_d-i+1)(i_M-i+2)}
      {(i_M-2i_x-2i_d+1)(i_M-i_d-i+2)} , \\
  & \frac{u_{i_xi_M+1i_d+1}(i)}{u_{i_xi_Mi_d}(i)} = \\
  & \hskip1em \frac{(i_d-i)(i_M-2i_x-2i_d)(i_M-i+2)}
     {(i_d+1)^2(i_x+i_d-i+1)} .
\end{aligned}\end{equation}
When~\eqref{eq:e,f} and~\eqref{eq:u/u} are
  inserted in~\eqref{eq:p+qr=as+bt}, it becomes
\begin{widetext}\begin{multline}\label{eq:veref}
  \frac{(i_M-i+2)(2j-i_M+i-1)}{(i_M+2)(2j-i_M-1)}
    + \frac{(2j-i+1)(i_d-i)(i_M-i_x-i_d-i+1)(i_M-i+2)}
        {(i_M+2)(2j-i_M-1)(i_x+i_d-i+1)(i_M-i_d-i+2)} = \\
    \frac{(i_M-2i_d+2)(2j-i_M+i_d-1)(i_M-i_x-i_d-i+1)(i_M-i+2)}
      {(i_M-i_x-2i_d+1)(i_M+2)(2j-i_M-1)(i_M-i_d-i+2)} \\
    + \frac{(2j-i_x-i_d)(i_d-i)(i_M-2i_x-2i_d)(i_M-i+2)}
        {(i_M+2)(i_M-i_x-2i_d+1)(2j-i_M-1)(i_x+i_d-i+1)} .
\end{multline}\end{widetext}
This relation holds identically, so~\eqref{eq:ef} is true for
$u_{i_xi_Mi_d}(i) \ne 0$ with $e _{i_xi_Mi_d}$ and
$f _{i_xi_Mi_d}$ given by~\eqref{eq:e,f} and can be shown to hold for
$u_{i_xi_Mi_d}(i) = 0$ by taking limits. Thus~\eqref{eq:ef} holds with
$e _{i_xi_Mi_d}$ and $f _{i_xi_Mi_d}$ given by~\eqref{eq:e,f}. The
omission of the second term in~\eqref{eq:ef} when $i_M$ is even and
$i_d = \lfloor i_M/2\rfloor-i_x$ does not invalidate this conclusion
because the second term on the right hand side of~\eqref{eq:veref}
vanishes in this case. Thus, since~\eqref{eq:ab=ef} is satisfied by
construction, \eqref{eq:tw=wt} is true, and, therefore, $V_{x_{i_x}}$
is rotationally invariant.

\begin{widetext}

\newpage\noindent

\section{\label{apx:j=9/2}Matrices $v(M)$ for
    $j = 9/2$}

\begin{equation}\begin{gathered}
v(8)=v(7)=v(-7)=v(-8)=\left(\begin{smallmatrix}
a
\end{smallmatrix}\right) , \\
v(6)=v(-6)^T=\left(\begin{smallmatrix}
a-b &
\tfrac{7}{3}b \\
b &
a-\tfrac{7}{3}b
\end{smallmatrix}\right) , \quad
v(5)=v(-5)^T=\left(\begin{smallmatrix}
a-2b &
\tfrac{8}{3}b \\
b &
a-\tfrac{4}{3}b
\end{smallmatrix}\right) , \\
v(4)=v(-4)^T=\left(\begin{smallmatrix}
a-3b+c &
\tfrac{25}{9}b-\tfrac{25}{9}c &
\tfrac{25}{6}c \\
b-c &
a-\tfrac{19}{9}b+\tfrac{25}{9}c &
\tfrac{8}{3}b-\tfrac{25}{6}c \\
c &
\tfrac{16}{9}b-\tfrac{25}{9}c &
a-4b+\tfrac{25}{6}c
\end{smallmatrix}\right) , \quad
v(3)=v(-3)^T=\left(\begin{smallmatrix}
a-4b+3c &
\tfrac{8}{3}b-\tfrac{16}{3}c &
\tfrac{25}{6}c \\
b-2c &
a-\tfrac{26}{9}b+\tfrac{32}{9}c &
\tfrac{25}{9}b-\tfrac{25}{9}c \\
c &
\tfrac{16}{9}b-\tfrac{16}{9}c &
a-\tfrac{20}{9}b+\tfrac{25}{18}c
\end{smallmatrix}\right) , \\
v(2)=v(-2)^T=\left(\begin{smallmatrix}
a-5b+6c-d &
\tfrac{7}{3}b-7c+\tfrac{7}{3}d &
\tfrac{7}{2}c-\tfrac{7}{2}d &
\tfrac{25}{6}d \\
b-3c+d &
a-\tfrac{11}{3}b+\tfrac{14}{3}c-\tfrac{7}{3}d &
\tfrac{8}{3}b-\tfrac{14}{3}c+\tfrac{7}{2}d &
\tfrac{25}{6}c-\tfrac{25}{6}d \\
c-d &
\tfrac{16}{9}b-\tfrac{28}{9}c+\tfrac{7}{3}d &
a-\tfrac{25}{9}b+\tfrac{49}{9}c-\tfrac{7}{2}d &
\tfrac{25}{9}b-\tfrac{125}{18}c+\tfrac{25}{6}d \\
d &
\tfrac{7}{3}c-\tfrac{7}{3}d &
\tfrac{7}{3}b-\tfrac{35}{6}c+\tfrac{7}{2}d &
a-5b+\tfrac{25}{3}c-\tfrac{25}{6}d
\end{smallmatrix}\right) , \\
v(1)=v(-1)^T=\left(\begin{smallmatrix}
a-6b+10c-4d &
\tfrac{16}{9}b-\tfrac{64}{9}c+\tfrac{16}{3}d &
\tfrac{7}{3}c-\tfrac{14}{3}d &
\tfrac{8}{3}d \\
b-4c+3d &
a-\tfrac{40}{9}b+\tfrac{55}{9}c-4d &
\tfrac{7}{3}b-\tfrac{35}{6}c+\tfrac{7}{2}d &
\tfrac{7}{2}c-2d \\
c-2d &
\tfrac{16}{9}b-\tfrac{40}{9}c+\tfrac{8}{3}d &
a-\tfrac{10}{3}b+\tfrac{35}{6}c-\tfrac{7}{3}d &
\tfrac{8}{3}b-4c+\tfrac{4}{3}d \\
d &
\tfrac{7}{3}c-\tfrac{4}{3}d &
\tfrac{7}{3}b-\tfrac{7}{2}c+\tfrac{7}{6}d &
a-\tfrac{8}{3}b+\tfrac{5}{2}c-\tfrac{2}{3}d
\end{smallmatrix}\right) , \\
v(0)=\left(\begin{smallmatrix}
a-7b+15c-10d+e &
b-5c+6d-e &
c-3d+e &
d-e &
e \\
b-5c+6d-e &
a-\tfrac{47}{9}b+\tfrac{71}{9}c-\tfrac{14}{3}d+e &
\tfrac{16}{9}b-\tfrac{52}{9}c+\tfrac{11}{3}d-e &
\tfrac{7}{3}c-3d+e &
\tfrac{8}{3}d-e \\
c-3d+e &
\tfrac{16}{9}b-\tfrac{52}{9}c+\tfrac{11}{3}d-e &
a-\tfrac{35}{9}b+\tfrac{109}{18}c-\tfrac{25}{6}d+e &
\tfrac{7}{3}b-\tfrac{14}{3}c+\tfrac{9}{2}d-e &
\tfrac{7}{2}c-\tfrac{14}{3}d+e \\
d-e &
\tfrac{7}{3}c-3d+e &
\tfrac{7}{3}b-\tfrac{14}{3}c+\tfrac{9}{2}d-e &
a-3b+\tfrac{13}{2}c-\tfrac{11}{2}d+e &
\tfrac{8}{3}b-\tfrac{15}{2}c+6d-e \\
e &
\tfrac{8}{3}d-e &
\tfrac{7}{2}c-\tfrac{14}{3}d+e &
\tfrac{8}{3}b-\tfrac{15}{2}c+6d-e &
a-\tfrac{16}{3}b+10c-\tfrac{20}{3}d+e
\end{smallmatrix}\right) .
\end{gathered}\end{equation}

\end{widetext}

\section{\label{apx:3f1110}Matrices $v(M)$ for
  $j = 9/2$ and $a,b,c,d,e = \frac34,1,1,1,0$}

\begin{equation}\begin{gathered}
v(8)=v(7)=v(-7)^T=v(-8)^T=\left(\begin{smallmatrix}
\tfrac{3}{4}
\end{smallmatrix}\right) , \\
v(6)=v(-6)^T=\left(\begin{smallmatrix}
-\tfrac{1}{4} &
\tfrac{7}{3} \\
1 &
-\tfrac{19}{12}
\end{smallmatrix}\right) , \\
v(5)=v(-5)^T=\left(\begin{smallmatrix}
-\tfrac{5}{4} &
\tfrac{8}{3} \\
1 &
-\tfrac{7}{12}
\end{smallmatrix}\right) , \\
v(4)=v(-4)^T=\left(\begin{smallmatrix}
-\tfrac{5}{4} &
0 &
\tfrac{25}{6} \\
0 &
\tfrac{17}{12} &
-\tfrac{3}{2} \\
1 &
-1 &
\tfrac{11}{12}
\end{smallmatrix}\right) , \\
v(3)=v(-3)^T=\left(\begin{smallmatrix}
-\tfrac{1}{4} &
-\tfrac{8}{3} &
\tfrac{25}{6} \\
-1 &
\tfrac{17}{12} &
0 \\
1 &
0 &
-\tfrac{1}{12}
\end{smallmatrix}\right) , \\
v(2)=v(-2)^T=\left(\begin{smallmatrix}
\tfrac{3}{4} &
-\tfrac{7}{3} &
0 &
\tfrac{25}{6} \\
-1 &
-\tfrac{7}{12} &
\tfrac{3}{2} &
0 \\
0 &
1 &
-\tfrac{1}{12} &
0 \\
1 &
0 &
0 &
-\tfrac{1}{12}
\end{smallmatrix}\right) , \\
v(1)=v(-1)^T=\left(\begin{smallmatrix}
\tfrac{3}{4} &
0 &
-\tfrac{7}{3} &
\tfrac{8}{3} \\
0 &
-\tfrac{19}{12} &
0 &
\tfrac{3}{2} \\
-1 &
0 &
\tfrac{11}{12} &
0 \\
1 &
1 &
0 &
-\tfrac{1}{12}
\end{smallmatrix}\right) , \\
v(0)=\left(\begin{smallmatrix}
-\tfrac{5}{4} &
2 &
-2 &
1 &
0 \\
2 &
-\tfrac{5}{4} &
-\tfrac{1}{3} &
-\tfrac{2}{3} &
\tfrac{8}{3} \\
-2 &
-\tfrac{1}{3} &
-\tfrac{5}{4} &
\tfrac{13}{6} &
-\tfrac{7}{6} \\
1 &
-\tfrac{2}{3} &
\tfrac{13}{6} &
-\tfrac{5}{4} &
\tfrac{7}{6} \\
0 &
\tfrac{8}{3} &
-\tfrac{7}{6} &
\tfrac{7}{6} &
-\tfrac{5}{4}
\end{smallmatrix}\right) .
\end{gathered}\end{equation}

\section{\label{apx:35t5h100}Matrices $v(M)$ for
  $j = 9/2$ and $a,b,c,d,e = \frac{35}{12},\frac52,1,1,0,0$}

\begin{equation}\begin{gathered}
v(8)=v(7)=v(-7)^T=v(-8)^T=\left(\begin{smallmatrix}
\tfrac{35}{12}
\end{smallmatrix}\right) , \\
v(6)=v(-6)^T=\left(\begin{smallmatrix}
\tfrac{5}{12} &
\tfrac{35}{6} \\
\tfrac{5}{2} &
-\tfrac{35}{12}
\end{smallmatrix}\right) , \\
v(5)=v(-5)^T=\left(\begin{smallmatrix}
-\tfrac{25}{12} &
\tfrac{20}{3} \\
\tfrac{5}{2} &
-\tfrac{5}{12}
\end{smallmatrix}\right) , \\
v(4)=v(-4)^T=\left(\begin{smallmatrix}
-\tfrac{43}{12} &
\tfrac{25}{6} &
\tfrac{25}{6} \\
\tfrac{3}{2} &
\tfrac{5}{12} &
\tfrac{5}{2} \\
1 &
\tfrac{5}{3} &
-\tfrac{35}{12}
\end{smallmatrix}\right) , \\
v(3)=v(-3)^T=\left(\begin{smallmatrix}
-\tfrac{49}{12} &
\tfrac{4}{3} &
\tfrac{25}{6} \\
\tfrac{1}{2} &
-\tfrac{3}{4} &
\tfrac{25}{6} \\
1 &
\tfrac{8}{3} &
-\tfrac{5}{4}
\end{smallmatrix}\right) , \\
v(2)=v(-2)^T=\left(\begin{smallmatrix}
-\tfrac{43}{12} &
-\tfrac{7}{6} &
\tfrac{7}{2} &
0 \\
-\tfrac{1}{2} &
-\tfrac{19}{12} &
2 &
\tfrac{25}{6} \\
1 &
\tfrac{4}{3} &
\tfrac{17}{12} &
0 \\
0 &
\tfrac{7}{3} &
0 &
-\tfrac{5}{4}
\end{smallmatrix}\right) , \\
v(1)=v(-1)^T=\left(\begin{smallmatrix}
-\tfrac{25}{12} &
-\tfrac{8}{3} &
\tfrac{7}{3} &
0 \\
-\tfrac{3}{2} &
-\tfrac{25}{12} &
0 &
\tfrac{7}{2} \\
1 &
0 &
\tfrac{5}{12} &
\tfrac{8}{3} \\
0 &
\tfrac{7}{3} &
\tfrac{7}{3} &
-\tfrac{5}{4}
\end{smallmatrix}\right) , \\
v(0)=\left(\begin{smallmatrix}
\tfrac{5}{12} &
-\tfrac{5}{2} &
1 &
0 &
0 \\
-\tfrac{5}{2} &
-\tfrac{9}{4} &
-\tfrac{4}{3} &
\tfrac{7}{3} &
0 \\
1 &
-\tfrac{4}{3} &
-\tfrac{3}{4} &
\tfrac{7}{6} &
\tfrac{7}{2} \\
0 &
\tfrac{7}{3} &
\tfrac{7}{6} &
\tfrac{23}{12} &
-\tfrac{5}{6} \\
0 &
0 &
\tfrac{7}{2} &
-\tfrac{5}{6} &
-\tfrac{5}{12}
\end{smallmatrix}\right) .
\end{gathered}\end{equation}

\bibliography{escuderos2}

\end{document}